\documentclass[fleqn,usenatbib]{mnras}%

\usepackage{newtxtext,newtxmath}

\usepackage[T1]{fontenc}

\DeclareRobustCommand{\VAN}[3]{#2}
\let\VANthebibliography\thebibliography
\def\thebibliography{\DeclareRobustCommand{\VAN}[3]{##3}\VANthebibliography}

\usepackage{graphicx}	
\usepackage{amsmath}	
\usepackage{booktabs}
\usepackage{lscape}
\usepackage{longtable}

\title[Circinus galaxy Inner Halo Globular Clusters]{Searching for globular clusters in the inner halo of the Circinus galaxy}

\author[C. Obasi]{
C. Obasi,$^{1,2}$\thanks{E-mail: casmiroluabuchukwuobasi@gmail.com}
M. G\'omez,$^{1}$
 D. Minniti$^{1,3,4}$
 L. D. Baravalle$^{5}$
M.V. Alonso$^{5,6}$
  B.I. Okere$^{2}$
\\
$^{1}$Instituto de Astrof\'isica, Facultad de Ciencias Exactas, Universidad Andres Bello,\\ Av. Fernandez Concha 700, Las Condes, Santiago, Chile.\\
$^{2}$Centre for Basic Space Science, University of Nigeria, Nsukka Nigeria.\\
$^{3}$Vatican Observatory, V00120 Vatican City State, Italy.\\
$^{4}$Departamento de Fisica, Universidade Federal de Santa Catarina, Trinidade 88040-900, Florianopolis, Brazil.\\
$^{5}$Instituto de Astronom\'ia Te\'orica y Experimental (IATE, CONICET-UNC),  Laprida 854, C\'ordoba, Argentina.\\
$^{6}$ Observatorio Astron\'omico de C\'ordoba, Universidad Nacional de C\'ordoba, Laprida 854, C\'ordoba, Argentina.
}

\date{Accepted}

\pubyear{2015}

\begin{document}
\label{firstpage}
\pagerange{\pageref{firstpage}--\pageref{lastpage}}
\maketitle

\begin{abstract}

 In this study, we search for Globular Clusters (GCs) in the inner halo of the Circinus galaxy using a combination of observational data. Our dataset includes observations from the VISTA Variables in the Vía Láctea Extended Survey (VVVX), optical data from Gaia Release 3 (DR3), and observations from the Dark Energy Camera (DECam). These multiple data sources provide a comprehensive basis for our analysis. Our search was concentrated within a 50 kpc radius from the centre, leading to the identification of 93 sources that met our established criteria. To ensure the reliability of our findings, we conducted multiple examinations for sample contamination. These examinations incorporated tests based on Gaia Astrometric Excess Noise (AEN), the Blue Photometer (BP)/Red Photometer (RP) Excess Factor (BRexcess), as well as comparisons with  stellar population models 
 This analysis confidently classified 41 sources as genuine GCs, as they successfully passed both the 3$\sigma$ Gaia AEN and BRexcess tests. We used the ISHAPE program to determine the structural parameters (half-light radii) of the GC candidates, with a peak effective radius of 4$\pm$ 0.5 pc. The catalogue mainly consists of bright GCs. Relationships between colour, size, and distance were found in the GC candidates, alongside confirmation of bi-modality in colour distributions.
\end{abstract}




\section{Introduction}

The presence of globular clusters (GCs) is common to almost all galaxies, regardless of luminosity or gas content. Due to their presence in nearly every galaxy, GCs provide us with a fundamental understanding of star formation, metal enrichment, and the merging histories of galaxies \citep{peng2006acs2}. Most studies \citep{kundu1998wide,kundu1999globular,puzia1999age,larsen2001g1,barmby2002m31,jordan2004possible,jordan2005acs,gomez2007sizes,forbes2010accreted,downing2012there,schulman2012effect} have shown that metal-rich GCs have half-light radii (r$_{e}$) that are 20$\%$ smaller than those of metal-poor GCs. Since this parameter remains relatively constant over GC lifetimes \citep{gomez2007sizes}, it serves as a good indicator of proto-GC sizes that are still observable today. A major constraint on the evolution of elliptical galaxies has been provided by the colour distributions of GC systems \citep{peng2006acs2}.  Colour-magnitude diagrams of
GCs in early-type galaxies have attracted considerable attention as a constraint on the formation of elliptical galaxies \citep{bower1992precision,peng2006acs2}.  While the mean colour of a galaxy can be estimated by analyzing its detailed star formation and chemical enrichment histories, GCs offer unique insights into galaxy evolution as they represent individual major epochs of star formation. GCs, being mostly single-age and single-metallicity systems, provide a valuable perspective on the evolutionary processes within galaxies. 

Compared to elliptical galaxies, our understanding of globular cluster (GC) systems in spiral and disk galaxies is still in its early stages. The bulk of our knowledge about GCs in these types of galaxies comes from studies of samples found in the Milky Way (MW) and M31, with only a limited number of investigations into more distant galaxies \citep[see][]{kissler1999hubble,chandar2004globular,strader2006globular,bridges2007spectroscopy,cantiello2007globular,degraaff2007galaxy,mora2007imaging,rhode2007global,harris2010diamonds,mayya2013nature,simanton2017gemini,wang2021properties,lomeli2022luminosity}. It is necessary to consider number statistics when evaluating the similarities and differences between spiral/disk galaxies and elliptical galaxies. Therefore, spiral/disk GC catalogue sample sizes need to be increased to draw any useful conclusions from the comparison. Additionally, to find the intrinsic nature of these GCs, it would be necessary to examine them closely across all types of host galaxies.

This paper presents a follow-up GC catalogue of the inner halo of the Circinus galaxy, which is a starburst and the nearest Seyfert II galaxy located about 4 Mpc away \citep{freeman1977large}. Due to its high foreground stellar density and significant dust obscuration, the Circinus galaxy is a challenging target for observation \citep{obasi2023globular}. Our GC catalogue, which combines our previous work with new data, will aid in designing appropriate spectroscopic follow-up observations to confirm the nature of the GC candidates.
The paper is organized as follows: Sect. \ref{sec2} describes the observations and data reduction process, while Sect. \ref{sec3} discusses the photometric pipeline used in this study. The method used for selecting candidate GCs is outlined in Sect. \ref{sec4}, and the results are presented in Sect. \ref{sec5}. Finally, we present our discussions and conclusions in Section \ref{sec6}.

\section{Observational data}\label{sec2}
In this analysis, we utilized images from the VISTA Variables in the Vía Láctea Extended survey \citep[VVVX,][]{minniti2018mapping} and data from Gaia Data Release 3 \citep[DR3,][]{Gaia2021}, which were previously described in \cite{obasi2023globular} (hereafter referred to as Paper 1), where the focus was on locating the Circinus galaxy outer halo GC population.  
This study utilized VVVX data comprising three NIR passbands (JHKs).  The observations were conducted under favourable conditions, with measured FWHM values of 1.0, 0.9, and 0.8 arcsec for the J, H, and K${_s}$ pass-bands, respectively. The total exposure time for the VISTA tile field amounted to 162 seconds. Additionally, we incorporated our DECam optical-view observations of the Circinus galaxy.  The DECam observations were performed using three optical filters (gri'). Details regarding DECam data processing can be found in the appendix \ref{app}. This multiwavelength approach, combining both NIR and optical data, serves as a robust method for studying GCs in high-density regions. 
  
Based on the catalogue of parameters for Galactic GCs by \cite{harris1996catalog}, it is typically observed that GCs have effective radii within the range of approximately 2 to 10 pc. This corresponds to angular sizes of 0.1 arcsec to 0.5 arcsec considering the distance (D=4 Mpc) to the Circinus galaxy, where 1 arcsec is $\sim$ 20 pc. Combining data from Gaia, with a resolution of 0.059 arcsec/pixel, and VVVX, with a resolution of 0.34 arcsec/pixel, a fraction of large GCs could be marginally resolved. The data utilized for this study consisted of science-ready images. Standard pipelines were used to reduce these images; then, we performed the photometry using the SExtractor program.
Our approach to  point source detection follows a similar procedure as in Paper 1, with a few differences that we describe below. 
\subsection{Unsharp Masking Technique}

\begin{figure*}
     \centering
         \centering
         \includegraphics[width=\linewidth]{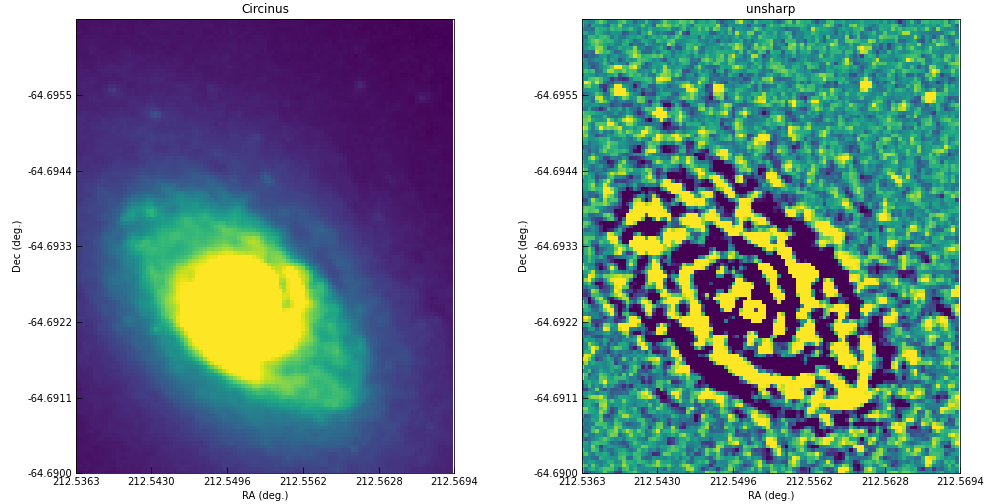}
       \caption{The distribution of the Circinus light profile in the K$_{s}$ passband is presented in the left panel. The unsharp mask image of Circinus is shown in the right panel. The unsharp mask image was convolved using a 2D Gaussian with a kernel size of 15x15. Discrete point sources are visible toward the center of the galaxy after applying the unsharp masking image techniques, allowing for the detection of faint sources.}
        \label{1}
\end{figure*}

There is a non-uniform extinction pattern near the Circinus galaxy centre, caused by 
the galaxy light profile (Paper 1). 
The distribution of the galaxy light profile makes it difficult to resolve distinct compact objects within the bulge and inner halo of the Circinus galaxy. To mitigate some of these challenges, we used the unsharp masking technique to remove the galaxy light profile. Unsharp masking is an image sharpening technique used to smooth the original image, enhancing the contrast of fine structure \citep{malin1979photo}. Compared with other galaxy modelling programmes, unsharp masking delivers better results in terms of the recovery of shrouded discrete point sources covered by the smooth light distribution of the galaxy \citep{buzzo2022new}. This is because, unlike in other modelling programmes like 
GALFIT 
\citep{peng2002detailed} and Isophote \citep{ciambur2015beyond}, where galaxy fitting methods are strongly dependent on the number of components fitted, and on how well the models describe the galaxy light profile, unsharp masking does not depend on any component to be fitted.  We utilized a 2D Gaussian kernel of size 15x15  pixels after testing various kernel sizes. The normalization of the kernel was set to false. Figure \ref{1} shows the K$_s$-band image, presenting both the smooth light distribution (left panel) and the discrete point source distribution obtained through unsharp masking of the original K$_s$-band image (right panel). 
Notably, the right panel reveals several discrete point sources, which allows us to detect fainter sources in those regions.

\begin{figure*}
     \centering
     
         \centering
         \includegraphics[width=\linewidth]{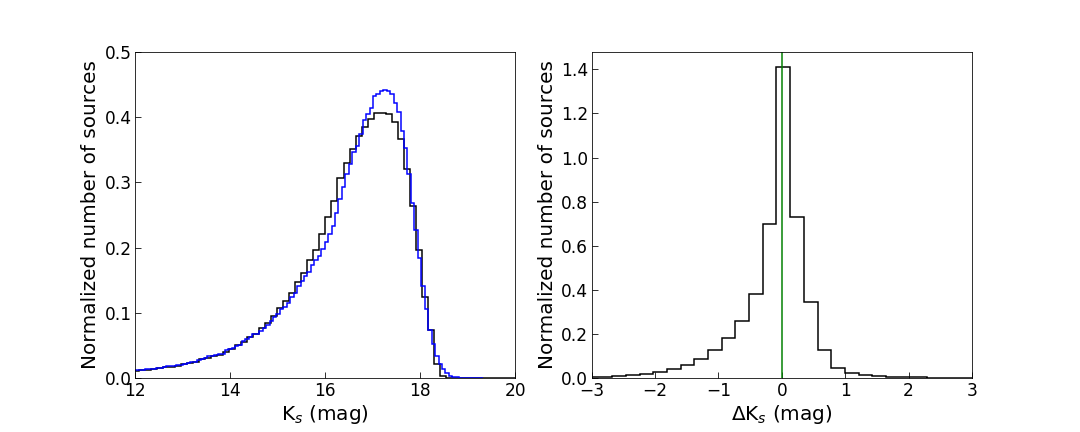}

       \caption{In the left panel, we compare the K$_{s}$ magnitude from our improved source detection with those from Paper 1, showing the black line for the latter and the blue line for the former.        In the right panel, we plot the mean difference ($\Delta$K$_{s}$) = K$_{s}$ \textsubscript{this work}-K$_{s}$ \textsubscript{Paper 1} in magnitudes.
       The green line shows the peak of the histogram.   
       }
        \label{a}
\end{figure*}

\section{The Photometric Pipeline}\label{sec3}

This section describes the procedure we used to detect and analyze point sources in the VVVX images, which is based on the combination of SExtractor v2.19.5 \citep{bertin1996SExtractor} and PSFEx v3.17.1 \citep{bertin2011asp}.

\subsection{PSFEx and SExtractor}
We created a point spread function (PSF) model for our photometry using the PSFEx software developed by \cite{bertin2011asp}. Initially, we ran SExtractor on a small region of the image to generate a catalogue of non-saturated stars, which served as input for PSFEx. Our PSF model was constructed as a linear combination of Gaussian or normal basis functions with specific criteria. For instance, we used a 20x20 pixel kernel, an elongation greater than 0.98, and a half-light radius that encloses 50$\%$ of the object's total flux. Our methodology for building the PSF model was similar to that outlined in \cite{baravalle2018searching} to search extragalactic candidates. 
We minimized the goodness of $\chi^2$ between the flux distribution and the model to obtain the PSF model. Finally, we applied the PSF model to the images and performed PSF photometry in SExtractor, with magnitudes estimated by integrating the sources over the model.

SExtractor was used to perform our photometry in Paper 1, where we discussed how its parameters allowed us to distinguish between point sources and extended sources. 
 
In this paper we have followed the photometric procedure described in detail by \cite{baravalle2018searching}.

Photometry was performed across the three JHK$_{s}$ selected filter bands using SExtractor in double image mode. Detection was performed with the K$_{s}$ unsharp image, while measurements were performed with the original JHK$_{s}$ images. 

In order to detect faint sources in the image, especially in the crowded region of the bulge, SExtractor configuration described in \cite{de2022j} was used:
BACK\_SIZE=4$\times$ FWHM$\times$1.05, BACK\_FILTERSIZE=2$\times$FWHM$\times$1.05 and PHOT\_AUTOPARAMS=3$\times$FWHM$\times$1.05.
By using these parameters, we were able to improve our ability to detect sources.

In summary, our photometric pipeline comprises multiple steps. First, we use SExtractor to create a catalogue that contains information such as  X and Y positions, the morphology of each object, flags, and an associated image. PSFEx then utilizes this catalogue to generate the best PSF model for each point source, searching for well-defined, circular, unsaturated, and isolated objects. Next, SExtractor applies this PSF model to extract the astrometric, photometric, and morphological properties of each source.

 To ensure the accuracy of our results, we compared our final PSF catalogue data with that presented in Paper 1 and found good agreement between both catalogues.  Figure \ref{a} shows the distribution of data from the two catalogues. The black line represents data from Paper 1, while the blue line represents data from this work, both in the K$_{s}$ passband (left panel). The right panel illustrates the difference between the two datasets, with a vertical green line indicating the centroid of the normalized differences.
 Our results show that we can detect fainter sources than with the standard SExtractor setting, as demonstrated in the left panel of Figure \ref{a}, where we compared the K$_{s}$ magnitude obtained with the standard SExtractor (black) with that obtained with the improved SExtractor configuration from \cite{de2022j} (blue). In comparing our previous work to the current study, it appears that we observed a higher number of sources at magnitudes K$_{s} =$ 15 and 16 mag. However, in this study, we detected a greater number of sources at fainter magnitudes, approximately around  K$_{s} =$ 17 mag. 
This comparison relies on normalized distributions. In the previous study, there were 1,665,767 sources detected in the K$_{s}$ band, while in this current work, we identified 1,674,949 sources. Consequently, we detected an additional 9182 new sources in this study, thanks to optimized settings that enable the detection of fainter sources.

 The concentration index of the GCs, defined as the ratio of aperture brightness at inner and outer radii, was also measured. This index is a crucial criterion for revealing the spatial extent and concentration of the GCs. For these measurements, we use SExtractor, which provides aperture magnitudes at different radii and total magnitudes.  In order to measure the concentration index to obtain quality GC candidates, we utilized aperture magnitudes within 6 and 12-pixel  radii, known for their effectiveness in distinguishing between point and extended sources \citep[see e.g;][]{peng2011hst,cantiello2018vegas}.
Aperture corrections were inferred from our previously published GC candidate catalogue in Paper 1, which served as a crucial quality control measure.  Utilizing the PSF photometry, we obtained total magnitudes (MAG\_AUTO) and aperture magnitudes at aperture radii of 6 and 12 pixels. Aperture corrections were computed by comparing MAG\_AUTO values from PSF photometry with magnitudes derived from these specific aperture sizes. For JHK${_s}$ bands, the determined aperture corrections were 0.73, 0.67, and 0.57 mag, respectively.

\subsection{Completeness of our photometry}
The completeness of our photometry in the VVV K$_{s}$ band has been extensively investigated in previous studies \citep[e.g.,][]{saito2012vvv,baravalle2018searching}. These studies consistently demonstrate that the K$_{s}$ band achieves a 90\% completeness level for magnitudes down to K$_{s} =$ 17.5 mag, and a 50\% completeness level for magnitudes down to  K$_{s} =$ 18 mag.
 
In our study, we specifically focus on the VVVX tiles located in the same region as Figure 11 of \cite{saito2012vvv}. We refer to tile b204, which has been thoroughly characterized for completeness. Additionally, in Paper 1, we briefly discussed the completeness of our data in JHK$_{s}$-bands, noting that both Paper 1 and our current work utilize the same dataset,
 and that the completeness does change slightly with the revised parameters.
Specifically, we found that our JHK$_{s}$ data has a 90\% completeness level at magnitudes 18.5, 18.0, and 17.5 respectively. 


\begin{figure}
     \centering
         \centering
         \includegraphics[width=\linewidth]{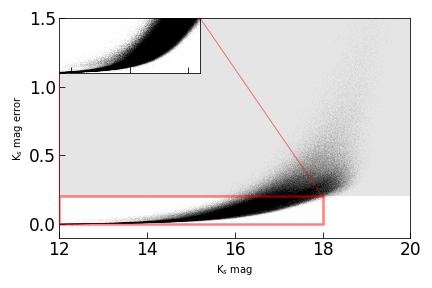}
        \caption{K$_{s}$ magnitudes vs K$_{s}$ magnitude errors of the sources detected in our photometry. The red rectangle shows reliable sources with measurement errors $\leq$ 0.2 mag that are of good quality. Sources with greater error measurements were discarded.
          }
        \label{2}
\end{figure}


\begin{figure*}
     \centering
         \centering
         \includegraphics[width=0.7\linewidth]{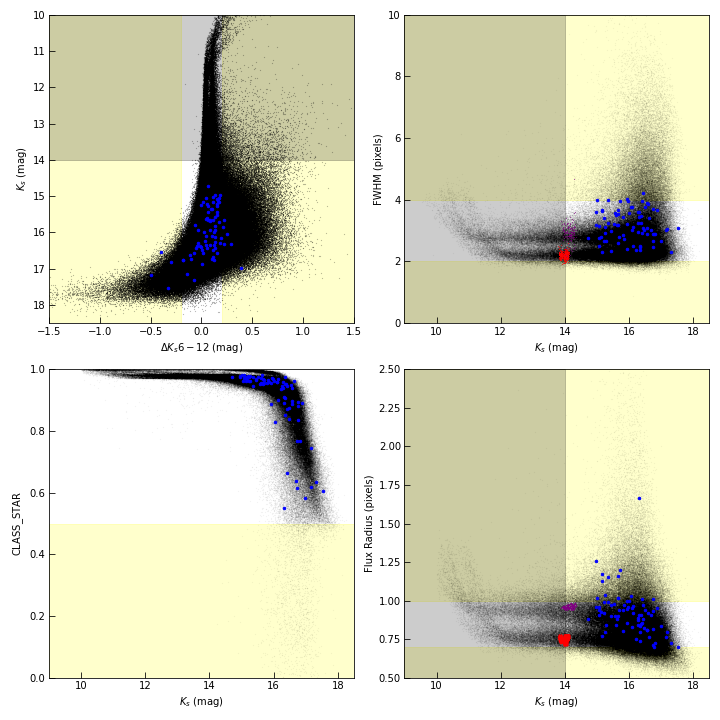}
        \caption{Our selection cuts. The upper panels show $\Delta$$_{X_{6-12}}$ versus K$_{s}$ magnitudes (left), and the K$_{s}$ magnitudes versus FWHM(right). The bottom panels show the K$_{s}$ magnitudes as a function of the CLASS\_STAR index (left), and the flux-radius (right).
The blue points represent our previously published GC catalogue. The shaded yellow regions have been excluded from our selection; only objects with K${_s}$ magnitudes < 14 mag were considered. On the left side of the upper panel, the two sequences correspond to the MW's faint main sequence and giant branch. The red and purple points represent a sample of stars from the colour-magnitude diagram of our data, illustrating the MW's faint main sequence and giant branch as foreground objects. The same explanation applies to the bottom panel.}
        \label{3}
\end{figure*}


\begin{figure*}
     \centering
         \centering
         \includegraphics[width=\linewidth]{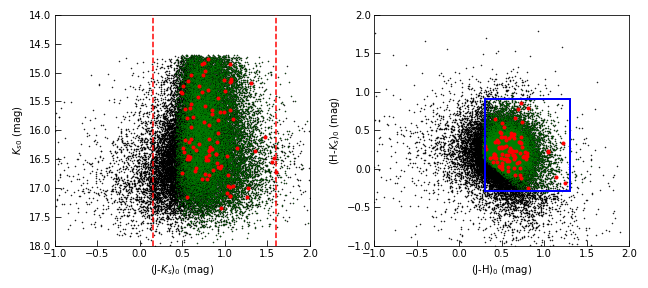}
        \caption{ Colour-magnitude diagram (left) and colour-colour diagram (right) of the point sources (black points). The green points represent our colour-colour selected point sources and the red points our final GC sample after applying the selection criteria. The dotted vertical red lines in the left panel represent the limits of the colour cut and the blue rectangle in the right panel indicates the visual location of our colour-colour selected point sources.
        }
        \label{5}
\end{figure*}

\begin{figure*}
     \centering
         \centering
        \includegraphics[width=\linewidth]{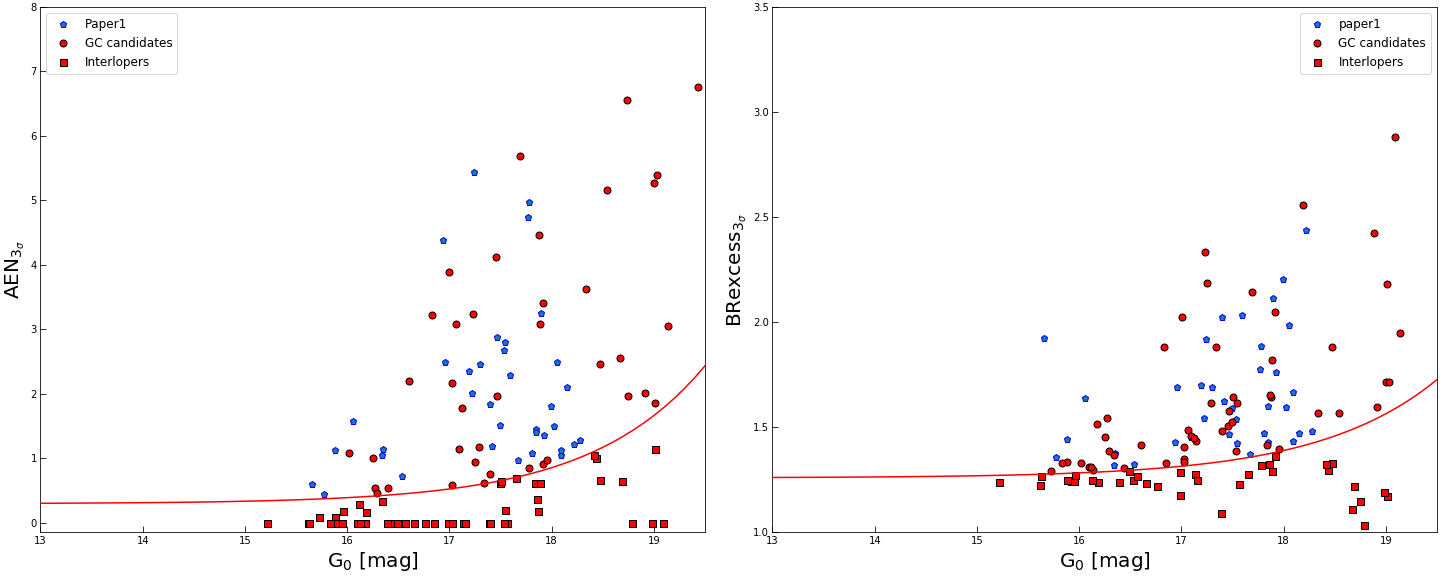}
        \caption{AEN and BRexcess are plotted against G magnitudes in the left and right panels, respectively. In this plot, our previously published data from Paper 1 is represented by blue points. Additionally, red points indicate GC candidates with a 3 $\sigma$ probability of being genuine, while red squares below the red curve represent candidates that are most likely interlopers. This curve represents the adopted relation \citep[taken from][]{hughes2021ngc}, which distinguishes whether a source is a point source, in this case, a GC, or an interloper.}
        \label{SA1}
\end{figure*}

\begin{figure*}
    \centering
    \includegraphics[width=\linewidth]{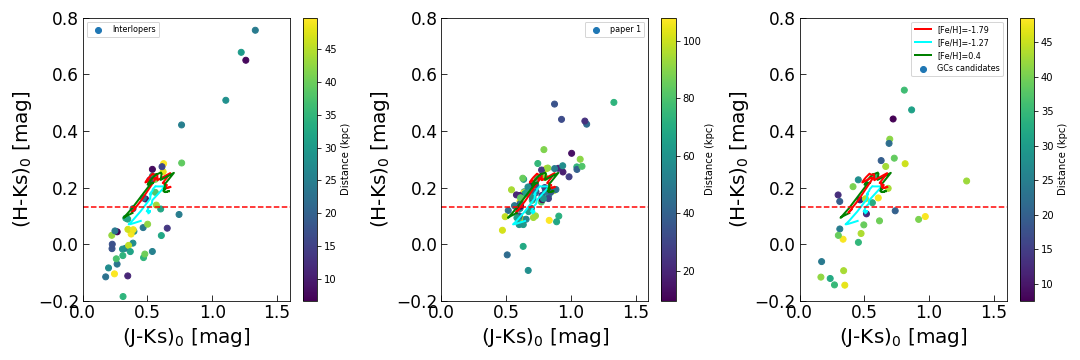}
    \caption{The colour-colour plots are presented with three representative integrated colour models of a single stellar population superimposed \citep[taken from][]{percival2008large}. In the left panel, we display the 52 targets identified as interlopers, while the middle panel showcases our previously published data from Paper 1. In the right panel, we present the 41 candidates that have a 3$\sigma$ probability of being genuine GCs. The dotted horizontal red line indicates the separation between objects that are 30 Myr or younger and those with ages spanning between 2-15 Gyrs. The colour bar denotes the distance from the galaxy centre, with dark blue and yellow points representing objects closer and farther, respectively. The legend displays the representative models and their [Fe/H] values. 
A notable observation is that the left panel is predominantly populated by younger objects, as indicated by the populations of objects below the dotted horizontal red line, in comparison to both the middle and right panels. }
    \label{SA2}
\end{figure*}

\begin{figure*}
    \centering
    \includegraphics[width=0.6\linewidth]{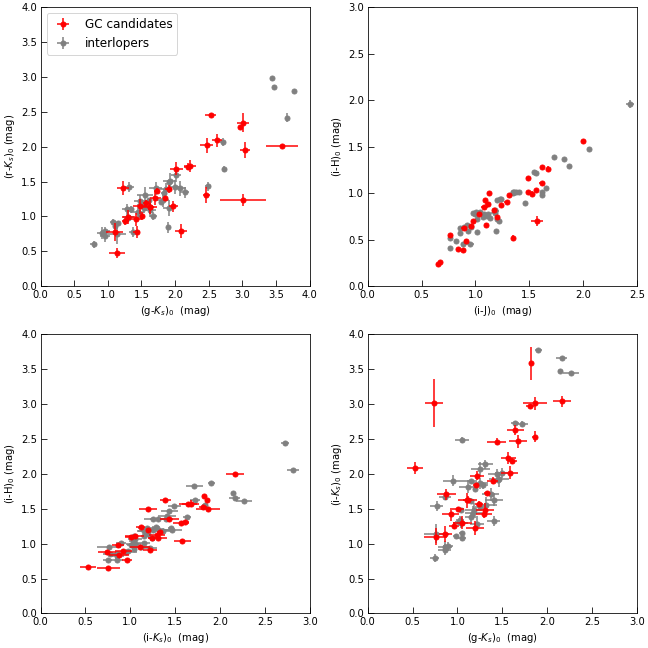}
    
    \caption{The optical/NIR colour-colour diagrams for the GC candidates and interlopers. In this representation, red points mark the 30 sources that have both VVVX and DECam magnitudes with a 3$\sigma$ probability of being genuine GCs, while gray points denote interlopers. Notably, our sample of interlopers exhibits a broader colour distribution compared to the 30 GC candidates due to their diverse stellar ages and masses. On average, these GC candidates appear one magnitude redder in colour space than the interlopers.
    }
    \label{fig:9a}
\end{figure*}

\begin{figure*}
     \centering
         \centering
         \includegraphics[width=\linewidth]{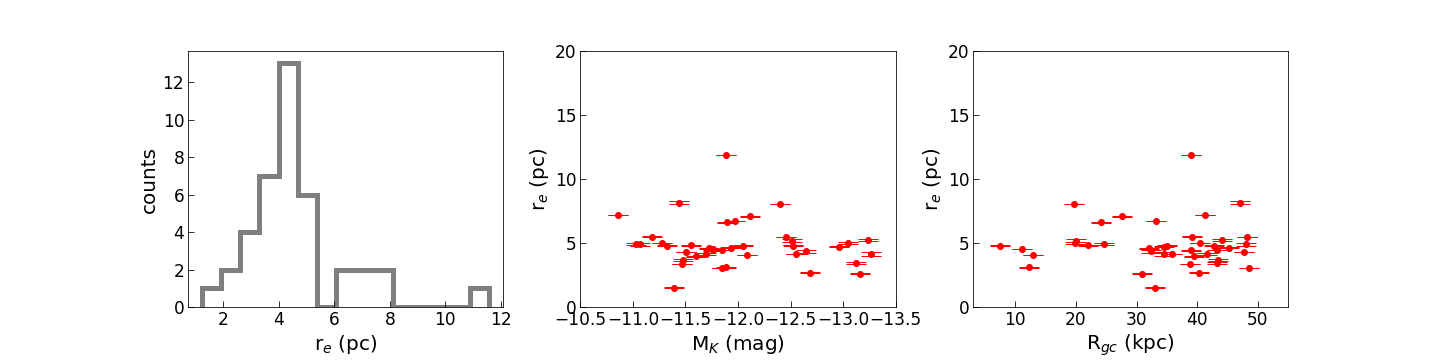}
        \caption{ The panels display the distribution of the effective radius (r$_e$) of the cluster candidates (left), the absolute luminosity of M$_{K_s}$ (middle), and the R$_{gc}$ (right) as functions of the effective radius. The effective radius peaks at 4$\pm$0.5 pc and the GC candidates are bright. In general, GC candidates farther away from the galaxy centre tend to have larger values of (r$_e$) compared to those closer to the centre.
        }
            \label{7}
\end{figure*}


\begin{figure*}
     \centering
         \centering
         \includegraphics[width=\linewidth]{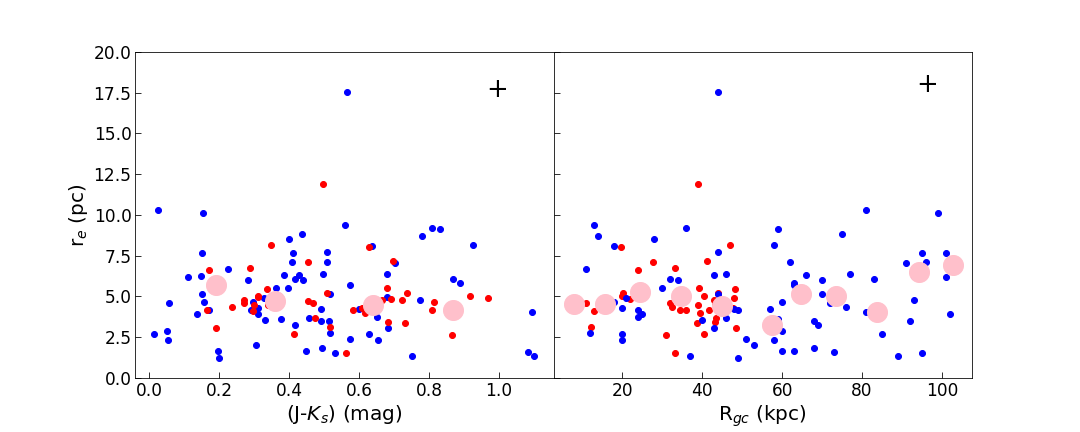}
        \caption{ Relationships between colour and distance with size are shown in the left and right panels, respectively. The pink points illustrate the median values of colour versus size and distance versus size. The plus sign shows the average error associated with the median values. Blue points represent our previously published catalogue, while red points represent the present study. 
        }
        \label{8}
\end{figure*}


\begin{figure*}
\includegraphics[width=\linewidth]{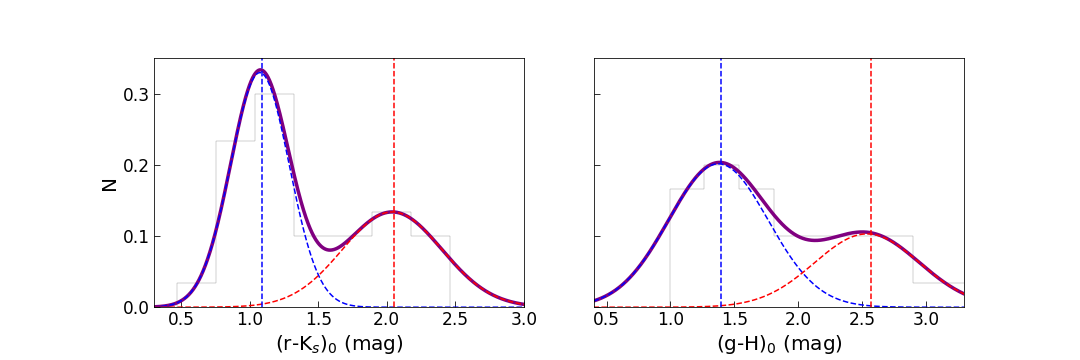}
\caption{ The optical/NIR (r-Ks)$_0$ and (g-H)$_0$ colour distribution for our GC candidates. The dashed blue and red curves show the distributions of the blue and red populations, while the solid curve represents the cumulative distribution. The dashed vertical lines represent the centroids for the blue and red sub-populations. The Gaussian fit reveals that the colour distributions are preferentially bimodal. N shows that the plot is normalised.
}
\label{9}
\end{figure*}


\begin{figure}
     \centering
     
         \centering
        \includegraphics[width=\linewidth]{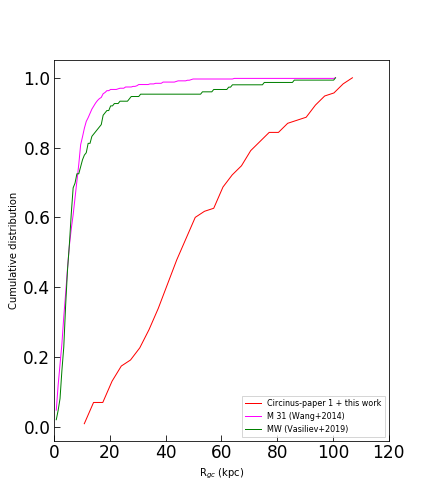}
     
        \caption{ The plot illustrates the cumulative distribution against the galactocentric distance of three GC catalogues: the Circinus galaxy, including Paper 1 from our previous study combined with those of this work, M31, and the MW, represented by red, pink, and green curves, respectively. The plot clearly shows that the spatial distribution of GCs in the Circinus galaxy is distinct from that in M31 and the MW.
        }
        \label{11}
\end{figure}


\begin{figure*}
\centering
         \centering
        \includegraphics[width=0.7\linewidth]{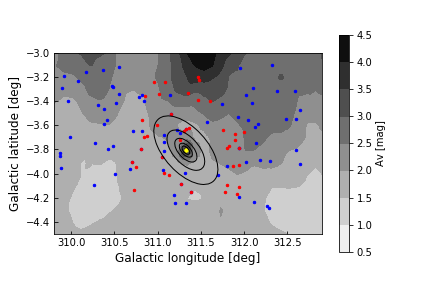}
        \caption{ The spatial distribution of GC candidates in the Circinus galaxy is separated into two populations: those identified in our previous study (Paper 1), represented by blue points, and those identified in this work, represented by red points. The yellow dot marks the adopted centre of the Circinus galaxy. The grayscale map shows the total optical interstellar extinctions, which were determined from the maps of \citep{schlafly2011measuringData}. The colour bar on the right indicates the extinction values. An ellipse, indicating the orientation of the galaxy, is superimposed, and there seems to be a lack of detection of GCs toward the inner disk region.
}
        \label{12}
\end{figure*}


\section{Selection of Globular Cluster Candidates}\label{sec4}
This study focuses on the inner halo of the Circinus galaxy, which has been historically challenging to observe due to high foreground extinction from the MW. The analysis primarily examines galactocentric distances less than 50 kpc, where a significant population of previously unidentified GCs is expected to reside.  In contrast, Paper 1 studied mostly the outer halo GCs of the Circinus galaxy.

 As a summary of the selection criteria used in Paper 1, the CLASS$\_$STAR index $\ge$ 0.5 was used to distinguish between point and extended sources, and colour cuts were applied to select sources that exhibit colours and magnitudes consistent with those expected for GCs in other galaxies. The mean and limits for the colours were as follows: 0.655, 0.113 $\leq$ (J-H)$_0$ $\leq$ 1.197 mag; 0.189, -0.424$\leq$ (H-K${s}$)$_0$ $\leq$ 0.802 mag; and 0.839, 0.159 $\leq$ (J-K${s}$)$_0$ $\leq$ 1.519 mag, respectively. To eliminate MW foreground stars, Gaia parameters (proper motion (pm) and parallax (plx)) information were used, while astrometric excess noise (AEN) and phot$\_$bp$\_$rp excess factor helped distinguish between point and extended sources for extragalactic objects. A spread model threshold of $\ge$0.002 was also employed to remove foreground objects. The morphology of the final sample was inspected using ellipticity ($\leq$0.4) and full width at half maximum (2 $\leq$ FWHM in pixel $\leq$ 4).

The photometric and morphometric selection criteria used in this study are similar to those in Paper 1. However, there are some differences that we explain below. The magnitude errors associated with our photometric measurements in the JHK${_s}$ passbands were examined, and only photometric measurements with errors smaller than 0.2 magnitudes were considered reliable for analysis as we show in Figure \ref{2}. Sources with SExtractor flag parameter greater than 4 were discarded in agreement with \cite{cho2016globular}. We measured the magnitude concentration index as described by \cite{cantiello2018vegas,peng2011hst}, which is the difference in magnitudes obtained within 6-pixel and 12-pixel aperture radii, 
($\Delta$ ${X_{6-12}}$, where X is the aperture corrected NIR magnitudes). For point sources, after applying the aperture correction, this difference should be zero. It is a powerful tool for identifying unresolved objects in distant galaxies, such as stars and GCs, that exhibit the properties of point-like sources.  We find that our VVVX dataset for the Circinus galaxy has excellent spatial resolution and FWHM in pixels, which allow some GCs to appear as slightly resolved extended sources in the dataset. As a result, the magnitude concentration index is close $\pm$ 0.2 for these objects.  Nonetheless, in general, the concentration index of GCs typically ranges from $\pm$0.1 to around $\pm$0.45, depending on the adopted aperture sizes used for the inner and outer radii.  
In comparison, the value for a foreground point source, such as stars, is expected to be around zero; however, they can easily be removed from the sample for their bright magnitudes. Similarly, the value for a point source, such as a star in a galaxy, also tends to be zero at similar faint magnitudes when compared with GCs, indicating their lower concentration \citep{hixenbaugh2022ancient}, which have values ranging from 0.1 to about 0.5. For galaxies, the values are even larger. Analysing the sample of the GC candidates presented in Paper 1, we found that $\Delta$ $_{X_{6-12}}$  for the majority of the GC candidates is $\sim$ $\pm$ 0.2 mag in the three NIR passbands consistent with other studies \citep[e.g;][]{peng2011hst,cantiello2018vegas}.   However, a few candidates in our previously published catalogue deviate from this distribution,
indicating that these candidates are most likely interlopers. We also used the SExtractor CLASS$\_$STAR parameter that
classifies objects into points and extended sources.
Following Paper 1, we discarded all of the sources with CLASS$\_$STAR values smaller than 0.5.
 
As a further refinement of the selection criteria for the K$_{s}$ passband, we utilized a broad set of SExtractor output parameters. These parameters included FWHM in pixels, flux radius in pixels (representing the radius within which half of the total source flux is contained, specifically for the GCs), ellipticity, and elongation.
These SExtractor selection criteria values mentioned above were determined through analysis of the parameters obtained in Paper 1. By comparing and evaluating these parameters, we were able to establish reliable and robust selection criteria for our investigation. 
Figure \ref{3} upper-left panel and upper-right panels show $\Delta_{X_{6-12}}$ as a function of K$_{s}$ magnitude and K$_{s}$ magnitude as a function of FWHM. In the lower panels, we show the K$_{s}$ magnitude as a function of CLASS$\_$STAR (left) and flux radius (right). 
  The blue points indicate our previously published GC catalogue while the shaded yellow regions are excluded from our selection. Our cut is restricted to Ks = 14 mag; any object brighter than this magnitude was removed as they are likely bright stars from the MW. Blue points located outside the unshaded regions are likely interlopers. 
The two sequences observed in the K$_{s}$ mag vs FWHM and K$_{s}$ mag vs Flux radius plots consist of stars belonging to the MW, including faint main sequences and faint giant stars. The red and purple points represent members of each sequence, respectively, selected from the colour-magnitude diagram of the entire dataset used for this study.

\begin{table}
\centering
\caption{The presented table summarises the selection criteria adopted for the analysis.}
\label{tab:my-table}
\resizebox{\columnwidth}{!}{
\begin{tabular}{@{}lll@{}}
\toprule
Quantity & Passband & Range adopted for selection \\ \midrule
\begin{tabular}[c]{@{}l@{}} Photometric  error\end{tabular}  & JHK${_s}$        & $\leq$ 0.2 mag                     \\
Flags parameter & JHK${_s}$        & $\leq$ 4                            \\
 $\Delta$$_{X_{6-12}}$                                        & K${_s}$          &  0 $\pm$ 0.1                           \\
CLASS$\_$STAR                                                 & K$_{s}$          & $\geq$ 0.5                            \\
FWHM (pixel)                                                  & K$_{s}$          &  2 $\leq$ (FWHM) (pixel) $\leq$  4            \\

Flux radius (pixel)                                 & K$_{s}$        & 0.6 $\leq$ r $\leq$1 \\
Ellipticity                                  & K$_{s}$        &    $\leq$ 0.4  \\
Elongation                                   & JHK$_{s}$      & 1  $\leq$ E $\leq$ 1.3   \\
Colour                                       & JHK$_{s}$      & 0.113 mag  $\leq$ (J-H)$_0$$\leq$ 1.197 mag\\
&                & -0.424 mag $\leq$ (H-K$_{s}$)$_0$ $\leq$ 0.802 mag\\ &  & 0.159 mag $\leq$ (J-K$_{s}$)$_0$ $\leq$ 1.519 mag
\\ \bottomrule
\end{tabular}%
}
\end{table}

As a final criterion in our selection process, we added colour cuts to minimise false detections from MW foreground stars. Magnitudes and colours were corrected for interstellar extinctions along the line of sight using \cite{schlafly2011measuringData} maps  that was
based on the \cite{fitzpatrick1999correcting} reddening law, 
and \cite{catelan2011vista} relative extinctions for VVV NIR passbands. These colour cuts were previously used in Paper 1. The range used was selected based on similar colour constraints used by \cite{jarrett20002mass} and \cite{baravalle2018searching} to select extragalactic sources (in their case, galaxies). MW foreground contamination posing as genuine GC objects is difficult to mitigate around the location of the Circinus galaxy. Possible sources of confusion include double stars, triple stars, and MW stellar associations with colours similar to GCs,
 
in addition to single stars with more uncertain concentration indices.

To remove these contaminants, we cross-matched our selected sample with the Gaia DR3 catalogue \citep{Gaia2021} and used two Gaia parameters pm and plx, following the same procedure as in Paper 1.  Gaia DR3 is crucial for eliminating MW foreground stars based on pms and plx values. Only objects with a proper motion probability value consistent with zero at 2$\sigma$ and a Plx probability within $\pm$0.4 are considered. Any source with higher values of pm or plx is excluded from the analysis.

Figure \ref{5} shows both the colour-magnitude diagram and the colour-colour diagram of our dataset. The black points represent the data, while the green points represent our applied colour cuts. Additionally, the red points indicate the decontaminated magnitudes and colours of our final sample, which can be observed in both the left and right panels. The red dotted lines on the left represent the colour limit. 
In Table \ref{tab:my-table}, we summarised our selection criteria.

\subsection{Contamination in Our Sample: Assessing Sources of Impurity}

In this study, the issue of galactic contamination presents a significant challenge that requires careful consideration. Hence, it is crucial to estimate and quantify the extent of contamination within the final sample. In our previous work (Paper 1), we employed the  Astrometric Excess Noise (AEN) and the blue photometer (BP)/Red Photometer (RP) Excess Factor (BRexcess) as selection criteria to mitigate galactic contamination. 
 The AEN and BRexcess factors are two Gaia parameters used to select objects slightly more extended than the PSF. AEN equals 0 for a well-fit star and increases for a poorer fit. BR excess is the ratio of the sum of fluxes in the blue and red photometers to the flux in the broad G passband. By combining the AEN and the BRexcess we can select GCs that are marginally more extended than the PSF. 
Utilizing the AEN and the BRexcess factor resolution, the \cite{hughes2021ngc} relation we adopted for this work can help us select GCs \citep{forbes2022low} and distinguish them from contaminants.
However, these criteria were optimized for larger radial distances from the galaxy center, proving more suitable for objects in the outer halo.
 
In the current study, our focus is on candidates within the inner halo of the galaxy. Although the majority of our final catalogue comprises candidates at larger distances, the AEN and BRexcess tests can still provide valuable insights into the level of contamination present. 
 Figure \ref{SA1} illustrates the results, depicting the AEN versus G magnitude on the left and BRexcess versus G magnitude on the right. The blue data points represent our previously published data, demonstrating a 3$\sigma$ probability of being GC candidates. The significance level of the 3$\sigma$ is solely associated with the AEN cut and is unrelated to the data points.  The red curve represents the adopted relation from \cite{hughes2021ngc}, defining the 3$\sigma$ criterion obtained through likelihood analysis. 
This likelihood analysis employs statistical distribution, specifically the Rayleigh distribution, to assess the likelihood of an object being classified as a foreground source based on its proper motion. This simple yet effective analysis enables us to identify and remove any source with significant proper motion, which is most likely a foreground object, thereby enhancing the overall reliability of the GC candidate sample. When we plot our data, we observe a clear separation into two distinct groups: the 41 red points located above the curve and the 52 red squares positioned below it. The red points exhibit a 3$\sigma$ probability of being genuine GCs, while the squares represent  interlopers.
 
Subsequently, we further investigate the nature of the contamination in our sample. We constructed a colour-colour plot and overlay it with  integrated colour models of a single stellar population from Basti\footnote{\url{http://basti.oa-abruzzo.inaf.it/BASTI/MAG_ML/intcolss.php}}, encompassing ages ranging from 30 Myr to 15 Gyr and varying metallicities \citep{percival2008large}. We selected representative models that match our colour-colour plots with [Fe/H] values of -1.79, -1.27, and 0.4 dex after testing several combinations. The left panel of Figure \ref{SA2} displays the colour-colour plot for the 52 candidates most likely interlopers, wherein we overlay the three representative integrated colour models. The middle panel showcases our previously published data from Paper 1. The right panel shows 41 candidates with a 3$\sigma$ probability of being genuine GCs. A dotted horizontal red line indicates the separation between objects that are 30 Myr or younger and those with ages spanning between 2 - 15 Gyrs. The legend in the left panel displays the [Fe/H] values of each model, while the colour bar indicates the distance from the galaxy centre, with dark blue and yellow points representing objects closer and farther, respectively. Objects below the curve in the left panel, with a mean J-K$_{s}$ colour of 0.75 mag, are likely younger, consistent with the study by \cite{baravalle2023agn}.
 The left panel of the plot corroborates the conclusions drawn in Figure \ref{SA1}, providing further evidence that objects positioned below the red curve are likely interlopers. Specifically, the mean J-K$_{s}$ colour for these objects is approximately 0.75 mag, in contrast to the candidates located above the curve. The latter group, with a 3$\sigma$ probability of being genuine GCs, exhibits a mean J-K$_{s}$ colour of 0.53 mag.   We estimate that our sample of 93 targets is contaminated by approximately 40\% to 50\% interlopers, as demonstrated in Figures \ref{SA1}-\ref{fig:9a}. The high contamination fraction in the sample underscores the challenges of studying GCs in highly spatially extincted fields, particularly the Circinus galaxy, which is veiled by foreground objects from the MW.
Consequently, for the remainder of our analysis, we will exclusively focus on the 41 sources that closely resemble GCs in our sample. A spectroscopic follow-up will be essential to further refine our dataset.

 \section{Results}\label{sec5}

From the total sample, we identified 93 targets. Among these, 41 sources passed a 3$\sigma$ Gaia AEN and BRexcess probability threshold, signifying their potential as genuine GC candidates. 
 We cross-referenced the data of the 41 GCs with DECam optical observations of the Circinus galaxy in the gri$'$ passbands using a matching radius of 0.5 arcsec.This merger resulted in 30 common sources between the two catalogues. This approach allowed us to create a final catalogue that includes different photometric data for use throughout the remainder of our analysis.

Figure \ref{fig:9a} presents the optical/NIR colour-colour diagrams of the GC candidates and interlopers that were cross-referenced with the DECam catalogue. In this representation, red points mark the 30 sources with a 3$\sigma$ probability of being genuine GCs, while gray points denote interlopers. Notably, our sample of interlopers exhibits a broader colour distribution compared to the smaller spread among the 30 GC candidates. We interpret this uniformity in colour distribution as indicative of a consistent age within our sample, a characteristic often associated with GCs \citep[see][for discussion]{puzia2004vlt}.

 When examining the colour-colour plots of the GC candidates and interlopers 
a general trend emerges. On average, the GC candidates appear one magnitude redder in colour space than the interlopers. However, the latter exhibits a broader range of colours due to their diverse stellar ages and masses.

Conversely, the GC is composed of older stars with a more uniform age and mass distribution. 
This contrast in colour distribution provides valuable insights into the age and mass characteristics of the GC candidates and interlopers.

\subsection{GC structural parameters}

The ISHAPE program \citep{larsen1999young} was utilised to determine the structural properties of each cluster candidate, specifically their effective or half-light radii (r$_{e}$). The K$_{s}$-band filter was chosen for this analysis due to its higher signal-to-noise ratio (S/N). For deriving the structural components of each cluster from the K$_{s}$ images, a model image was convolved with a point spread function (PSF) using ISHAPE.

Creating an accurate model that replicates the effects of telescope optics necessitates an accurate stellar PSF. To achieve this, we used the pstelect and PSF functions in DAOPHOT within IRAF to generate PSFs for each K$_{s}$ field using several bright, isolated stars. The empirical PSFs were spatially sub-sampled by a factor of 10, which is a requirement for the optimal performance of the ISHAPE program.

For simplicity, we assumed a KING profile shape, which is well-suited for a broad range of GCs found in various environments \citep{gomez2007sizes}. While we also attempted other analytical models, such as Moffat functions, these did not yield improvements in fitting. Potential systematic effects on sizes caused by different models were discussed by \cite{larsen1999young}. For sources with a similar extension to the stellar PSF, the effective radius appears to be independent of the model. The concentration index is defined as c=r$_{t}$/r$_{c}$ based on the profiles by \cite{king1962structure}, where r$_{t}$ represents the tidal radius of the cluster and r$_{c}$ denotes the core radius. In our VVVX image, we were able to fit parameters including ellipticity, position angle, and r$_{c}$.
 The values determined for a given cluster with varying concentration parameters did not differ by more than 10\% on average for the GCs.

The left panel of Figure~\ref{7} displays the effective radii distribution of the cluster candidates, with r$_{e}$ peaking at 4$\pm$0.5 pc, consistent with the values of GCs found in the MW \citep{harris1991globular} and NGC 5128 \citep{gomez2007sizes}. The middle panel illustrates the absolute luminosity of M$_K{_s}$ versus size r${_e}$. The results demonstrate that we are currently only capable of identifying the most luminous GC candidates within the Circinus galaxy. A plot of galactocentric distance R$_{gc}$ against r$_{e}$ is presented in the right panel. 
Objects located at greater galactocentric distances, on average, tend to have larger sizes than those situated at smaller galactocentric distances. This finding provides further evidence for the GC nature of our sample, as this result is consistent with findings in other galaxies \citep{barmby2002m31,jordan2005acs,gomez2007sizes}.

To explore whether the observed trends in cluster size and distance are genuine or a result of potential data incompleteness, we combined our previously published outer halo GC candidate catalogue for the Circinus galaxy with our inner halo data. We organised our data into bins of 0.25-magnitude colour and 10 kpc galactocentric distance, then calculated the median size for each bin. In Figure~\ref{8}, the left panel depicts r$_e$ vs. (J-Ks)$_0$ colour, and the right panel shows r$_e$ vs. R$_{gc}$, both overlaid with the median values of their corresponding colour and size.  
This plot confirms the trend that there is a clear relationship between GC size and colour, as well as size and galactocentric distance. This relationship further strengthens the argument for the GC nature of our sample, given its consistency with findings in GCs from other galaxies \citep{kundu1998wide,kundu1999globular,puzia1999age,larsen2001g1,barmby2002m31,jordan2005acs,gomez2007sizes}.

\begin{table*}
\caption{GMM test results}
\label{table2}
\resizebox{1.2\columnwidth}{!}{%
\begin{tabular}{@{}lllllllll@{}}
\toprule
Colour & N  & Blue           & Red           & N$_{blue}$          & N$_{red}$          & Kurt & DD            & p($\chi$$^2$) \\ \midrule
(r-K${_s}$)    & 30 & 1.09$\pm$0.05  & 2.05$\pm$0.11 & 20.5$\pm$2.9 & 9.5$\pm$2.9  & 0.25 & 3.93$\pm$0.80 & 0.026               \\
(i-K${_s}$)    & 30 & 1.05$\pm$0.02  & 1.71$\pm$0.12 & 18.1$\pm$4.4 & 11.9$\pm$4.4 & 0.27 & 3.02$\pm$0.70 & 0.415               \\
(g-H)    & 30 & 1.40$\pm$0.11  & 2.57$\pm$0.21 & 20.6$\pm$3.3 & 9.4$\pm$3.3  & 0.4  & 3.5$\pm$0.80  & 0.02                \\
(i-H)    & 30 & 1.027$\pm$0.05 & 1.58$\pm$0.11 & 19.6$\pm$4.4 & 10.4$\pm$4.4 & 0.3  & 2.97$\pm$0.86 & 0.31                \\ \bottomrule
\end{tabular}%
}
\end{table*}
\subsection{GC colour bimodality}
Table \ref{table2} presents the results of the bi-modality test conducted using the Gaussian mixture model \citep[GMM,][]{muratov2010modeling}. The analysis focused exclusively on the 30 GC candidates with measurements in both NIR VVVX and Optical DECam data. For each colour combination, we calculated the distribution peaks, fractions of blue and red populations, kurtosis, peak separation, and the $\chi^2$ value.

Conventionally, the red population is expected to be more prominent closer to the galactic centre compared to the blue population. However, our findings in the Circinus galaxy reveal a predominantly blue GC population, as summarised in Table~\ref{table2}. This intriguing outcome might be influenced by potential biases in our relatively shallow photometric data, which is insufficiently deep to reach the GC turnover point within the Circinus galaxy due to significant spatial extinction.

On the other hand, the situation can also be viewed as intrinsic; as noted by \cite{peng2006acs2}, fractions of the red GC population decrease as galaxy luminosity becomes fainter. Figure 6 in \cite{peng2006acs2}  shows the diminishing red GC population with decreasing galaxy luminosity, with an almost complete absence within the range -16 mag < M$_{B}$ < -15 mag. The blue GC population dominates in the range -20 mag < M$_{B}$ < -15 mag and decreases as the galaxies become brighter, specifically in the range -22 mag < M$_{B}$ < -20  mag. This result aligns with Figure 22 from \cite{harris2016globular}, depicting the metallicity distribution of MW GCs. Like many other galaxies with fainter magnitudes less than -20 mag, blue GCs dominate. The trend observed in the Circinus galaxy GCs appears to align with this pattern.

Nevertheless, Figure ~\ref{9} illustrates that our GC candidates are well-characterised by a bimodal distribution—a hallmark feature of GCs. To explore fainter magnitudes and reach the GC turnover in the Circinus galaxy, deeper NIR photometry within the range of K$_{s}$= 25–28 mag would be necessary, a capability provided by the James Webb Space Telescope \citep{vikaeus2022conditions}.  Additionally, we have explored how bimodality changes with radius, but as of now, no definitive conclusion has been reached. Further investigation requires a deeper photometric sample, and a new imaging proposal is currently in progress.

 \subsection{Radial and spatial distribution of the GCs}
Figure ~\ref{11} provides an updated comparison of the projected radial distances of GCs in the Circinus galaxy, M31, and the MW. The cumulative distribution against galactocentric distance is shown for three GC catalogues, representing the Circinus galaxy (including Paper 1 from our previous study combined with those of this work), M31, and the MW, with red, pink, and green curves, respectively. Our earlier study in Paper 1 showed significant GC concentrations for distances (D) $<$ 20 kpc in both M31 and the MW. For this revised analysis, we combined our previous data with 41 new GC candidates and plotted the distribution.
However, there still exists a lack of GC concentration at D $<$ 20 kpc in the Circinus galaxy, unlike the observed patterns in M31 and the MW. Assuming that the intrinsic radial distribution in the Circinus galaxy mirrors that of the MW and M31, our findings suggest the presence of approximately nine times more GC candidates within the inner 20 kpc compared to beyond this range. Notably, this estimate considerably exceeds the value of 120 GCs we previously predicted for this galaxy in Paper 1. Furthermore, a specific frequency (S$\_$N) at such a high level would deviate from the trends observed in other spiral galaxies. Another perspective to address this divergence is to consider that the MW's GCS might still be incomplete in its inner regions \citep{minniti2017new}. 

The distribution of GCs in Circinus suggests that this galaxy has an extended GC system, resembling those of the MW and M31.

In Figure~\ref{12}, we present the spatial distribution of our previous catalogue (represented by blue points) and our new catalogue represented by red points, all plotted in Galactic coordinates. The gray-scale bar indicates various A$_{V}$ levels from the optical maps by \cite{schlafly2011measuringData}. In addition, we have included an ellipse to show the orientation of the galaxy. It is worth noting that the detection of GC candidates increases as we move outward with an increase in galactocentric distance. Conversely, the detection rate significantly decreases inward, almost reaching zero at D=7 kpc. This observation appears to be consistent with the findings of \cite{gonzalez2022relation}, who reported a lack of GC candidate detections in the inner region of the spiral galaxies they studied using ground-based instruments.

 In Table~\ref{Table3}, we provide a machine-readable format listing all the GC candidates. Column (1) includes the identification and J2000 equatorial coordinates, while Columns (2) and (3) provide the J, H, and K$_{s}$ magnitudes, along with their respective errors. The ellipticity and A$_{V}$ values are presented in Columns (10) and (11), respectively, while Columns (12) and (13) present the galactocentric distances and effective radii.
\section{Discussion and Final remarks}\label{sec6}
In this study, we conducted an investigation into the inner halo of the Circinus galaxy. Among the 93 cluster candidates we identified, 41 sources exhibited a high likelihood of being genuine GC candidates, surpassing a 3$\sigma$ Gaia AEN and BRexcess probability threshold. We subsequently examined various properties of these 41 sources and compared them with GCs from other galaxies.

Our analysis of optical/NIR colour-colour diagrams revealed a distinctive absence of colour variation, in contrast to the broader colour distribution observed among the interlopers. This lack of colour variation suggests a uniform age within our sample, a characteristic commonly associated with GCs \citep{puzia2004vlt}.

Additionally, the effective radii of our cluster candidates, peaking at 4.0$\pm$0.5 pc, closely resemble values observed in GC populations within the MW \citep{harris1991globular} and NGC 5128 \citep{gomez2007sizes}. It is worth noting that our candidate clusters primarily comprise bright and blue GC candidates, possibly due to limitations in our photometric data.  
It can also be interpreted as an intrinsic property of GCs found in galaxies with similar luminosity to the MW, as noted by \cite{peng2006acs2}  and \cite{harris2016globular}. The fraction of the red GC population decreases with a decrease in galaxy luminosity. 

Our observations also unveiled distinct colour-size and distance-size relationships in the candidate clusters, a phenomenon observed in both galactic and extra-galactic GCs \citep{kundu1999globular,puzia1999age,larsen2001g1,barmby2002m31,jordan2005acs,gomez2007sizes}.

Moreover, the outcomes of the colour bi-modality test, employing the GMM test by \cite{muratov2010modeling}, indicate that our sample aligns well with a bimodal distribution, a characteristic often associated with GCs \citep{brodie2006extragalactic,brodie2014sages,harris2017globular}. Notably, blue GCs appear to predominate over red GCs in our sample, which might be influenced by potential photometric biases.

The cumulative radii distribution reveals distinctions between the distribution of the Circinus galaxy GCs and those in M31 and the MW. Taken together, these tests and evidence strongly support the classification of our sample as GC candidates. To further validate our findings and explore the fainter end of the Circinus galaxy GC luminosity function, NIR spectroscopy and deeper photometry \citep[e.g., data from the James Webb Space Telescope;][]{vikaeus2022conditions} would be required.

\subsection{Final remarks: The Origin of the Circinus galaxy Globular Cluster Population}

We can formulate a plausible hypothesis regarding the origin of the Circinus galaxy GCs.  Our proposition is grounded in well-established principles and observations regarding the formation and evolution of galaxy haloes and is supported by the results of our investigations. As established in numerous studies \citep{bullock2005tracing,abadi2006stars,font2006chemical,forbes2010accreted}, the accretion of smaller galaxies containing dark matter, stars, gas, and GCs is the primary process through which MW-like galaxy haloes are built. During this process, smaller galaxies are often disrupted and accreted by larger galaxies, with the high density of their GCs enabling them to survive and contribute to the pre-existing GC population of the host galaxy. This process is further complemented by the in situ formation of stars and clusters within the host galaxy. Our findings indicate that the majority of the Circinus galaxy GC candidates  are located beyond the 20 kpc which is where most of the material is expected to be accreted \cite[See ][for discussion]{forbes2010accreted}. This is highly suggestive of an accretion-dominated formation scenario, in which the GCs of smaller, accreted galaxies have contributed significantly to the GC population of the Circinus galaxy halo. Furthermore, our results indicate that there is likely a significant population of in situ GCs in the Circinus galaxy that have yet to be fully explored due to limitations in our current photometric data. 

To further validate our hypothesis, we anticipate detecting a number of stellar streams located between 200 to 500 kpc from the Circinus galaxy, which would serve as compelling evidence of the accretion and disruption processes that transpired \cite{helmi1999debris,newberg2002ghost,duffau2005spectroscopy,grillmair2009four,thomas2020hidden,ibata2021charting,martin2022pristine,miro2023search}. Such evidence would strongly support our proposal that the Circinus galaxy GC was predominantly formed through accretion and that the in situ contribution to the GC population of the host galaxy is likely limited.

Our findings provide insights into the formation history and evolution of the Circinus galaxy and further underscore the importance of continued investigations into the origins of the Circinus galaxy globular clusters.

\begin{table}
\caption{Near-IR VVVX photometry, colour and shape parameters of our cluster candidates.}
\label{Table3}
\resizebox{\columnwidth}{!}{%
\begin{tabular}{@{}llllllllllllll@{}}
\toprule
ID   & RA       & DEC      & J     & $\sigma$J & H     & $\sigma$H & K$_s$ & $\sigma$K$_s$ & FWHM & $\varepsilon$ & A$_V$ & R$_{gc}$ & r$_{e}$ \\ \midrule

1.0   & 211.859  & -65.463  & 16.26 & 0.03 & 15.83 & 0.05 & 15.81 & 0.12 & 3.06 & 0.2  & 1.88 & 22 & 5.9   \\
2.0   & 211.9072 & -65.0491 & 16.89 & 0.03 & 16.36 & 0.04 & 16.07 & 0.1  & 3.74 & 0.26 & 2.01 & 45 & 4.64  \\
3.0   & 211.9313 & -65.6224 & 15.28 & 0.01 & 14.94 & 0.02 & 14.78 & 0.04 & 3.17 & 0.29 & 1.64 & 44 & 5.23  \\

4.0   & 211.9691 & -65.2543 & 16.55 & 0.04 & 16.28 & 0.05 & 16.25 & 0.11 & 3.88 & 0.26 & 1.69 & 39 & 4.46  \\

5.0   & 212.1935 & -65.3633 & 16.56 & 0.04 & 15.92 & 0.05 & 16.29 & 0.15 & 3.04 & 0.27 & 1.75 & 31 & 4.58  \\
6.0   & 212.0443 & -64.9166 & 16.92 & 0.07 & 16.43 & 0.09 & 16.57 & 0.16 & 3.62 & 0.21 & 2.35 & 47 & 8.14  \\
7.0   & 212.3727 & -65.246  & 16.35 & 0.03 & 16.12 & 0.06 & 15.89 & 0.14 & 3.58 & 0.3  & 2.24 & 27 & 7.08  \\
8.0   & 212.2574 & -64.9891 & 16.62 & 0.05 & 16.19 & 0.08 & 16.12 & 0.14 & 3.91 & 0.24 & 2.48 & 38 & 11.87 \\
9.0   & 212.7034 & -65.4799 & 16.24 & 0.02 & 15.91 & 0.03 & 15.61 & 0.06 & 3.45 & 0.31 & 1.77 & 19 & 8.06  \\
10   & 212.8541 & -65.5965 & 17.15 & 0.05 & 16.82 & 0.05 & 16.46 & 0.14 & 2.77 & 0.2  & 1.6  & 21 & 4.82  \\
11   & 212.3405 & -64.874  & 16.16 & 0.02 & 15.77 & 0.03 & 15.49 & 0.09 & 3.83 & 0.22 & 2.69 & 42 & 4.8   \\
12   & 212.9879 & -65.5976 & 16.24 & 0.03 & 15.62 & 0.04 & 15.5  & 0.13 & 2.86 & 0.29 & 1.58 & 20 & 5.18  \\
13   & 212.693  & -65.103  & 16.28 & 0.04 & 16.05 & 0.04 & 16.11 & 0.13 & 3.26 & 0.26 & 2.18 & 24 & 6.62  \\
14   & 212.1427 & -65.8321 & 17.11 & 0.06 & 16.67 & 0.09 & 16.5  & 0.17 & 3.77 & 0.33 & 1.61 & 47 & 4.29  \\
15   & 212.0701 & -65.6453 & 17.85 & 0.07 & 17.52 & 0.09 & 17.15 & 0.2  & 3.47 & 0.33 & 1.63 & 41 & 7.19  \\
16   & 212.1072 & -65.4834 & 16.27 & 0.02 & 16.0  & 0.04 & 15.46 & 0.11 & 3.2  & 0.25 & 1.6  & 35 & 4.15  \\
17   & 212.0921 & -65.38   & 16.24 & 0.02 & 15.82 & 0.03 & 15.97 & 0.11 & 3.01 & 0.2  & 1.66 & 35 & 4.79  \\
18  & 213.0685 & -65.284  & 17.41 & 0.08 & 17.13 & 0.09 & 16.69 & 0.19 & 3.28 & 0.21 & 2.34 & 7  & 4.81  \\
19  & 213.3721 & -65.621  & 15.28 & 0.02 & 15.12 & 0.02 & 14.97 & 0.05 & 3.63 & 0.21 & 1.58 & 19 & 5.01  \\
20  & 213.138  & -65.1949 & 16.71 & 0.04 & 16.4  & 0.03 & 16.24 & 0.13 & 2.73 & 0.3  & 1.92 & 11 & 4.57  \\
21  & 213.1648 & -65.1714 & 16.64 & 0.04 & 16.23 & 0.04 & 16.13 & 0.15 & 2.36 & 0.22 & 1.88 & 12 & 3.13  \\
22  & 213.0117 & -64.8797 & 17.18 & 0.09 & 16.76 & 0.09 & 16.62 & 0.13 & 2.58 & 0.21 & 3.09 & 33 & 1.53  \\
23  & 213.2444 & -65.155  & 16.22 & 0.04 & 16.1  & 0.05 & 15.92 & 0.12 & 3.77 & 0.28 & 1.9  & 12 & 4.09  \\
24  & 213.6954 & -65.6493 & 17.29 & 0.06 & 17.03 & 0.05 & 16.98 & 0.14 & 3.01 & 0.25 & 1.5  & 24 & 4.96  \\
25  & 213.1609 & -64.7225 & 16.5  & 0.05 & 16.07 & 0.05 & 16.16 & 0.1  & 3.29 & 0.25 & 3.83 & 43 & 4.46  \\
26  & 213.3099 & -64.8966 & 15.72 & 0.02 & 15.33 & 0.02 & 14.85 & 0.06 & 2.96 & 0.23 & 2.78 & 30 & 2.63  \\
27 & 213.2074 & -64.7431 & 16.48 & 0.07 & 16.2  & 0.06 & 16.31 & 0.14 & 3.81 & 0.27 & 3.74 & 41 & 4.16  \\
28 & 213.6258 & -64.8646 & 15.33 & 0.02& 14.99 & 0.02& 14.75 & 0.05& 2.59 & 0.27& 3.02    & 34  & 4.16    \\
29 & 214.1385 & -65.0403 & 15.6  & 0.02 & 15.24 & 0.03 & 15.36 & 0.07 & 3.73 & 0.26 & 2.19 & 32 & 4.38  \\
30 & 214.5826 & -65.5178 & 17.02 & 0.04 & 16.49 & 0.04 & 16.41 & 0.13 & 3.12 & 0.2  & 1.9  & 39 & 3.99  \\
31   & 214.3518 & -65.1694 & 16.33 & 0.03 & 15.81 & 0.04 & 16.04 & 0.08 & 3.1  & 0.27 & 2.11 & 33 & 6.73  \\
32   & 214.5934 & -65.457  & 17.28 & 0.05 & 16.85 & 0.05 & 16.55 & 0.1  & 2.91 & 0.23 & 1.93 & 38 & 3.36  \\
33   & 214.3838 & -65.1511 & 15.51 & 0.01 & 15.06 & 0.01 & 15.05 & 0.04 & 2.6  & 0.21 & 2.12 & 34 & 4.73  \\
34   & 214.6257 & -65.284  & 17.51 & 0.06 & 16.58 & 0.06 & 16.83 & 0.13 & 3.69 & 0.24 & 2.12 & 39 & 5.52  \\
35   & 214.4571 & -65.029  & 17.66 & 0.11 & 16.82 & 0.08 & 16.74 & 0.18 & 3.2  & 0.23 & 2.22 & 40 & 5.01  \\
36   & 214.513  & -65.0723 & 15.74 & 0.02 & 15.53 & 0.02 & 15.33 & 0.07 & 2.47 & 0.22 & 2.2  & 40 & 2.71  \\
37   & 214.9094 & -65.4894 & 15.89 & 0.02 & 15.57 & 0.02 & 15.55 & 0.07 & 2.81 & 0.28 & 1.8  & 48 & 5.46  \\
38   & 214.7671 & -65.2587 & 17.01 & 0.05 & 16.57 & 0.04 & 16.53 & 0.15 & 2.59 & 0.21 & 2.12 & 43 & 3.7   \\
39   & 214.9299 & -65.4264 & 17.92 & 0.12 & 17.04 & 0.12 & 16.95 & 0.17 & 3.23 & 0.31 & 1.81 & 48 & 4.91  \\
40   & 214.676  & -65.1239 & 15.57 & 0.02 & 15.08 & 0.02 & 14.88 & 0.05 & 3.68 & 0.27 & 2.3  & 43 & 3.46  \\
41   & 214.6987 & -64.9745 & 16.36 & 0.04 & 15.87 & 0.04 & 16.17 & 0.12 & 2.78 & 0.27 & 2.51 & 48 & 3.05  \\

 \bottomrule
\end{tabular}%
}
\end{table}

\section*{Acknowledgements}

We gratefully acknowledge the use of data from the ESO Public Survey program IDs 179.B-2002 and 198.B-2004 taken with the VISTA telescope and data products from the Cambridge Astronomical Survey Unit.

This work has made use of data from the European Space Agency (ESA) mission
{\it Gaia} (\url{https://www.cosmos.esa.int/gaia}), processed by the {\it Gaia}
Data Processing and Analysis Consortium (DPAC,
\url{https://www.cosmos.esa.int/web/gaia/dpac/consortium}). Funding for the DPAC has been provided by national institutions, in particular, the institutions participating in the {\it Gaia} Multilateral Agreement. We acknowledge the use of data obtained with the Dark Energy Camera (DECam), mounted on the Blanco 4-meter telescope at the Cerro Tololo Inter-American Observatory (CTIO) in Chile.  We also acknowledge the comments of the anonymous reviewer whose positive feedback helped to improve the quality of this paper. 
 D.M. gratefully acknowledges support by the ANID BASAL projects ACE210002 and FB210003, and from CNPq/Brazil through project 350104/2022-0. 
D.M. and M. G. also gratefully acknowledge support from Fondecyt Project No. 1220724.
 
\section*{Data Availability}

The VVVX data used in this article can be accessed through the ESO science portal via the following link: \url{https://archive.eso.org/scienceportal/home}. The tile names, as provided in the article, should be used for retrieval. Gaia data can be accessed at \url{https://gea.esac.esa.int/archive/}. However, please note that the DECam data is subject to proprietary restrictions at this moment.



\bibliographystyle{mnras}
\bibliography{main} 

\begin{thebibliography}{}
\makeatletter
\relax
\def\mn@urlcharsother{\let\do\@makeother \do\$\do\&\do\#\do\^\do\_\do\%\do\~}
\def\mn@doi{\begingroup\mn@urlcharsother \@ifnextchar [ {\mn@doi@} {\mn@doi@[]}}
\def\mn@doi@[#1]#2{\def\@tempa{#1}\ifx\@tempa\@empty \href {http://dx.doi.org/#2} {doi:#2}\else \href {http://dx.doi.org/#2} {#1}\fi \endgroup}
\def\mn@eprint#1#2{\mn@eprint@#1:#2::\@nil}
\def\mn@eprint@arXiv#1{\href {http://arxiv.org/abs/#1} {{\tt arXiv:#1}}}
\def\mn@eprint@dblp#1{\href {http://dblp.uni-trier.de/rec/bibtex/#1.xml} {dblp:#1}}
\def\mn@eprint@#1:#2:#3:#4\@nil{\def\@tempa {#1}\def\@tempb {#2}\def\@tempc {#3}\ifx \@tempc \@empty \let \@tempc \@tempb \let \@tempb \@tempa \fi \ifx \@tempb \@empty \def\@tempb {arXiv}\fi \@ifundefined {mn@eprint@\@tempb}{\@tempb:\@tempc}{\expandafter \expandafter \csname mn@eprint@\@tempb\endcsname \expandafter{\@tempc}}}

\bibitem[\protect\citeauthoryear{Abadi, Navarro  \& Steinmetz}{Abadi et~al.}{2006}]{abadi2006stars}
Abadi M.~G.,  Navarro J.~F.,   Steinmetz M.,  2006, MNRAS, 365, 747

\bibitem[\protect\citeauthoryear{Baravalle, Alonso, Castell{\'o}n, Beam{\'\i}n  \& Minniti}{Baravalle et~al.}{2018}]{baravalle2018searching}
Baravalle L.~D.,  Alonso M.~V.,  Castell{\'o}n J. L.~N.,  Beam{\'\i}n J.~C.,   Minniti D.,  2018, AJ, 155, 46

\bibitem[\protect\citeauthoryear{Baravalle et~al.,}{Baravalle et~al.}{2023}]{baravalle2023agn}
Baravalle L.~D.,  et~al., 2023, MNRAS, 520, 5950

\bibitem[\protect\citeauthoryear{Barmby, Holland  \& Huchra}{Barmby et~al.}{2002}]{barmby2002m31}
Barmby P.,  Holland S.,   Huchra J.~P.,  2002, AJ, 123, 1937

\bibitem[\protect\citeauthoryear{Bertin \& Arnouts}{Bertin \& Arnouts}{1996}]{bertin1996SExtractor}
Bertin E.,  Arnouts S.,  1996, A\&A, 117, 393

\bibitem[\protect\citeauthoryear{Bertin, Evans, Accomazzi, Mink  \& Rots}{Bertin et~al.}{2011}]{bertin2011asp}
Bertin E.,  Evans I.,  Accomazzi A.,  Mink D.,   Rots A.,  2011, ASP Conf. Ser. Vol. 442, Astronomical Data Analysis Software and Systems XX

\bibitem[\protect\citeauthoryear{Bower, Lucey  \& Ellis}{Bower et~al.}{1992}]{bower1992precision}
Bower R.~G.,  Lucey J.,   Ellis R.~S.,  1992, MNRAS, 254, 589

\bibitem[\protect\citeauthoryear{Bridges, Rhode, Zepf  \& Freeman}{Bridges et~al.}{2007}]{bridges2007spectroscopy}
Bridges T.~J.,  Rhode K.~L.,  Zepf S.~E.,   Freeman K.~C.,  2007, AJ, 658, 980

\bibitem[\protect\citeauthoryear{Brodie \& Strader}{Brodie \& Strader}{2006}]{brodie2006extragalactic}
Brodie J.~P.,  Strader J.,  2006, ARA\&A, 44, 193

\bibitem[\protect\citeauthoryear{Brodie et~al.,}{Brodie et~al.}{2014}]{brodie2014sages}
Brodie J.~P.,  et~al., 2014, AJ, 796, 52

\bibitem[\protect\citeauthoryear{Bullock \& Johnston}{Bullock \& Johnston}{2005}]{bullock2005tracing}
Bullock J.~S.,  Johnston K.~V.,  2005, AJ, 635, 931

\bibitem[\protect\citeauthoryear{Buzzo et~al.,}{Buzzo et~al.}{2022}]{buzzo2022new}
Buzzo M.~L.,  et~al., 2022, MNRAS, 510, 1383

\bibitem[\protect\citeauthoryear{Cantiello, Blakeslee  \& Raimondo}{Cantiello et~al.}{2007}]{cantiello2007globular}
Cantiello M.,  Blakeslee J.~P.,   Raimondo G.,  2007, AJ, 668, 209

\bibitem[\protect\citeauthoryear{Cantiello et~al.,}{Cantiello et~al.}{2018}]{cantiello2018vegas}
Cantiello M.,  et~al., 2018, A\&A, 611, A93

\bibitem[\protect\citeauthoryear{Catelan et~al.,}{Catelan et~al.}{2011}]{catelan2011vista}
Catelan M.,  et~al., 2011, in RR Lyrae Stars, Metal-Poor Stars, and the Galaxy.

\bibitem[\protect\citeauthoryear{Chandar, Whitmore  \& Lee}{Chandar et~al.}{2004}]{chandar2004globular}
Chandar R.,  Whitmore B.,   Lee M.~G.,  2004, AJ, 611, 220

\bibitem[\protect\citeauthoryear{Cho, Blakeslee, Chies-Santos, Jee, Jensen, Peng  \& Lee}{Cho et~al.}{2016}]{cho2016globular}
Cho H.,  Blakeslee J.~P.,  Chies-Santos A.~L.,  Jee M.~J.,  Jensen J.~B.,  Peng E.~W.,   Lee Y.-W.,  2016, AJ, 822, 95

\bibitem[\protect\citeauthoryear{Ciambur}{Ciambur}{2015}]{ciambur2015beyond}
Ciambur B.~C.,  2015, AJ, 810, 120

\bibitem[\protect\citeauthoryear{DeGraaff, Blakeslee, Meurer  \& Putman}{DeGraaff et~al.}{2007}]{degraaff2007galaxy}
DeGraaff R.~B.,  Blakeslee J.~P.,  Meurer G.~R.,   Putman M.~E.,  2007, AJ, 671, 1624

\bibitem[\protect\citeauthoryear{Downing}{Downing}{2012}]{downing2012there}
Downing J.,  2012, MNRAS, 425, 2234

\bibitem[\protect\citeauthoryear{Duffau, Zinn, Vivas, Carraro, M{\'e}ndez, Winnick  \& Gallart}{Duffau et~al.}{2005}]{duffau2005spectroscopy}
Duffau S.,  Zinn R.,  Vivas A.~K.,  Carraro G.,  M{\'e}ndez R.~A.,  Winnick R.,   Gallart C.,  2005, AJ, 636, L97

\bibitem[\protect\citeauthoryear{Fitzpatrick}{Fitzpatrick}{1999}]{fitzpatrick1999correcting}
Fitzpatrick E.~L.,  1999, PASP, 111, 63

\bibitem[\protect\citeauthoryear{Font, Johnston, Bullock  \& Robertson}{Font et~al.}{2006}]{font2006chemical}
Font A.~S.,  Johnston K.~V.,  Bullock J.~S.,   Robertson B.~E.,  2006, AJ, 638, 585

\bibitem[\protect\citeauthoryear{Forbes \& Bridges}{Forbes \& Bridges}{2010}]{forbes2010accreted}
Forbes D.~A.,  Bridges T.,  2010, MNRAS, 404, 1203

\bibitem[\protect\citeauthoryear{Forbes, Ferr{\'e}-Mateu, Gannon, Romanowsky, Carlin, Brodie  \& Day}{Forbes et~al.}{2022}]{forbes2022low}
Forbes D.~A.,  Ferr{\'e}-Mateu A.,  Gannon J.~S.,  Romanowsky A.~J.,  Carlin J.~L.,  Brodie J.~P.,   Day J.,  2022, MNRAS, 512, 802

\bibitem[\protect\citeauthoryear{Freeman et~al.,}{Freeman et~al.}{1977}]{freeman1977large}
Freeman K.,  et~al., 1977, A\&A, 55, 445

\bibitem[\protect\citeauthoryear{{Gaia Collaboration} et~al.,}{{Gaia Collaboration} et~al.}{2021}]{Gaia2021}
{Gaia Collaboration} et~al., 2021, \mn@doi [\aap] {10.1051/0004-6361/202039657}, \href {https://ui.adsabs.harvard.edu/abs/2021A&A...649A...1G} {649, A1}

\bibitem[\protect\citeauthoryear{G{\'o}mez \& Woodley}{G{\'o}mez \& Woodley}{2007}]{gomez2007sizes}
G{\'o}mez M.,  Woodley K.~A.,  2007, AJ, 670, L105

\bibitem[\protect\citeauthoryear{Gonz{\'a}lez-L{\'o}pezlira et~al.,}{Gonz{\'a}lez-L{\'o}pezlira et~al.}{2022}]{gonzalez2022relation}
Gonz{\'a}lez-L{\'o}pezlira R.~A.,  et~al., 2022, ApJ, 941, 53

\bibitem[\protect\citeauthoryear{Grillmair}{Grillmair}{2009}]{grillmair2009four}
Grillmair C.~J.,  2009, AJ, 693, 1118

\bibitem[\protect\citeauthoryear{Harris}{Harris}{1991}]{harris1991globular}
Harris W.~E.,  1991, ARA\&A, 29, 543

\bibitem[\protect\citeauthoryear{Harris}{Harris}{1996}]{harris1996catalog}
Harris W.~E.,  1996, Astronomical Journal v. 112, p. 1487, 112, 1487

\bibitem[\protect\citeauthoryear{Harris, Spitler, Forbes  \& Bailin}{Harris et~al.}{2010}]{harris2010diamonds}
Harris W.~E.,  Spitler L.~R.,  Forbes D.~A.,   Bailin J.,  2010, MNRAS, 401, 1965

\bibitem[\protect\citeauthoryear{Harris, Blakeslee, Whitmore, Gnedin, Geisler  \& Rothberg}{Harris et~al.}{2016}]{harris2016globular}
Harris W.~E.,  Blakeslee J.~P.,  Whitmore B.~C.,  Gnedin O.~Y.,  Geisler D.,   Rothberg B.,  2016, ApJ, 817, 58

\bibitem[\protect\citeauthoryear{Harris, Ciccone, Eadie, Gnedin, Geisler, Rothberg  \& Bailin}{Harris et~al.}{2017}]{harris2017globular}
Harris W.~E.,  Ciccone S.~M.,  Eadie G.~M.,  Gnedin O.~Y.,  Geisler D.,  Rothberg B.,   Bailin J.,  2017, AJ, 835, 101

\bibitem[\protect\citeauthoryear{Helmi, White, De~Zeeuw  \& Zhao}{Helmi et~al.}{1999}]{helmi1999debris}
Helmi A.,  White S.~D.,  De~Zeeuw P.~T.,   Zhao H.,  1999, Nature, 402, 53

\bibitem[\protect\citeauthoryear{Hixenbaugh, Chandar  \& Mok}{Hixenbaugh et~al.}{2022}]{hixenbaugh2022ancient}
Hixenbaugh K.,  Chandar R.,   Mok A.,  2022, AJ, 163, 271

\bibitem[\protect\citeauthoryear{Hughes et~al.,}{Hughes et~al.}{2021}]{hughes2021ngc}
Hughes A.~K.,  et~al., 2021, APJ, 914, 16

\bibitem[\protect\citeauthoryear{Ibata et~al.,}{Ibata et~al.}{2021}]{ibata2021charting}
Ibata R.,  et~al., 2021, AJ, 914, 123

\bibitem[\protect\citeauthoryear{Jarrett, Chester, Cutri, Schneider, Skrutskie  \& Huchra}{Jarrett et~al.}{2000}]{jarrett20002mass}
Jarrett T.,  Chester T.,  Cutri R.,  Schneider S.,  Skrutskie M.,   Huchra J.,  2000, AJ, 119, 2498

\bibitem[\protect\citeauthoryear{Jord{\'a}n}{Jord{\'a}n}{2004}]{jordan2004possible}
Jord{\'a}n A.,  2004, AJ, 613, L117

\bibitem[\protect\citeauthoryear{Jord{\'a}n et~al.,}{Jord{\'a}n et~al.}{2005}]{jordan2005acs}
Jord{\'a}n A.,  et~al., 2005, AJ, 634, 1002

\bibitem[\protect\citeauthoryear{King}{King}{1962}]{king1962structure}
King I.,  1962, AJ, 67, 471

\bibitem[\protect\citeauthoryear{Kissler-Patig, Ashman, Zepf  \& Freeman}{Kissler-Patig et~al.}{1999}]{kissler1999hubble}
Kissler-Patig M.,  Ashman K.~M.,  Zepf S.~E.,   Freeman K.~C.,  1999, AJ, 118, 197

\bibitem[\protect\citeauthoryear{Kundu \& Whitmore}{Kundu \& Whitmore}{1998}]{kundu1998wide}
Kundu A.,  Whitmore B.~C.,  1998, AJ, 116, 2841

\bibitem[\protect\citeauthoryear{Kundu, Whitmore, Sparks, Macchetto, Zepf  \& Ashman}{Kundu et~al.}{1999}]{kundu1999globular}
Kundu A.,  Whitmore B.~C.,  Sparks W.~B.,  Macchetto F.~D.,  Zepf S.~E.,   Ashman K.~M.,  1999, AJ, 513, 733

\bibitem[\protect\citeauthoryear{Larsen}{Larsen}{1999}]{larsen1999young}
Larsen S.,  1999, A\&AS, 139, 393

\bibitem[\protect\citeauthoryear{Larsen}{Larsen}{2001}]{larsen2001g1}
Larsen S.~S.,  2001, AJ, 122, 1782

\bibitem[\protect\citeauthoryear{Lomel{\'\i}-N{\'u}{\~n}ez, Mayya, Rodr{\'\i}guez-Merino, Ovando  \& Rosa-Gonz{\'a}lez}{Lomel{\'\i}-N{\'u}{\~n}ez et~al.}{2022}]{lomeli2022luminosity}
Lomel{\'\i}-N{\'u}{\~n}ez L.,  Mayya Y.,  Rodr{\'\i}guez-Merino L.,  Ovando P.,   Rosa-Gonz{\'a}lez D.,  2022, MNRAS, 509, 180

\bibitem[\protect\citeauthoryear{Malin}{Malin}{1979}]{malin1979photo}
Malin D.,  1979, Mercury, 8, 89

\bibitem[\protect\citeauthoryear{Martin et~al.,}{Martin et~al.}{2022}]{martin2022pristine}
Martin N.~F.,  et~al., 2022, MNRAS, 516, 5331

\bibitem[\protect\citeauthoryear{Mayya, Rosa-Gonz{\'a}lez, Santiago-Cort{\'e}s, Rodr{\'\i}guez-Merino, Vega, Torres-Papaqui, Bressan  \& Carrasco}{Mayya et~al.}{2013}]{mayya2013nature}
Mayya Y.~D.,  Rosa-Gonz{\'a}lez D.,  Santiago-Cort{\'e}s M.,  Rodr{\'\i}guez-Merino L.,  Vega O.,  Torres-Papaqui J.,  Bressan A.,   Carrasco L.,  2013, MNRAS, 436, 2763

\bibitem[\protect\citeauthoryear{Minniti}{Minniti}{2018}]{minniti2018mapping}
Minniti D.,  2018, in , The Vatican Observatory, Castel Gandolfo: 80th Anniversary Celebration.
Springer, pp 63--71

\bibitem[\protect\citeauthoryear{Minniti et~al.,}{Minniti et~al.}{2017}]{minniti2017new}
Minniti D.,  et~al., 2017, ApJL, 849, L24

\bibitem[\protect\citeauthoryear{Mir{\'o}-Carretero et~al.,}{Mir{\'o}-Carretero et~al.}{2023}]{miro2023search}
Mir{\'o}-Carretero J.,  et~al., 2023, A\&A, 669, L13

\bibitem[\protect\citeauthoryear{Mora, Larsen  \& Kissler-Patig}{Mora et~al.}{2007}]{mora2007imaging}
Mora M.~D.,  Larsen S.~S.,   Kissler-Patig M.,  2007, A\&A, 464, 495

\bibitem[\protect\citeauthoryear{Muratov \& Gnedin}{Muratov \& Gnedin}{2010}]{muratov2010modeling}
Muratov A.~L.,  Gnedin O.~Y.,  2010, AJ, 718, 1266

\bibitem[\protect\citeauthoryear{Newberg et~al.,}{Newberg et~al.}{2002}]{newberg2002ghost}
Newberg H.~J.,  et~al., 2002, AJ, 569, 245

\bibitem[\protect\citeauthoryear{Obasi et~al.,}{Obasi et~al.}{2023}]{obasi2023globular}
Obasi C.,  et~al., 2023, A\&A, 670, A18

\bibitem[\protect\citeauthoryear{Peng, Ho, Impey  \& Rix}{Peng et~al.}{2002}]{peng2002detailed}
Peng C.~Y.,  Ho L.~C.,  Impey C.~D.,   Rix H.-W.,  2002, AJ, 124, 266

\bibitem[\protect\citeauthoryear{Peng et~al.,}{Peng et~al.}{2006}]{peng2006acs2}
Peng E.~W.,  et~al., 2006, AJ, 639, 95

\bibitem[\protect\citeauthoryear{Peng et~al.,}{Peng et~al.}{2011}]{peng2011hst}
Peng E.~W.,  et~al., 2011, AJ, 730, 23

\bibitem[\protect\citeauthoryear{Percival, Salaris, Cassisi  \& Pietrinferni}{Percival et~al.}{2008}]{percival2008large}
Percival S.~M.,  Salaris M.,  Cassisi S.,   Pietrinferni A.,  2008, AJ, 690, 427

\bibitem[\protect\citeauthoryear{Puzia, Kissler-Patig, Brodie  \& Huchra}{Puzia et~al.}{1999}]{puzia1999age}
Puzia T.~H.,  Kissler-Patig M.,  Brodie J.~P.,   Huchra J.~P.,  1999, AJ, 118, 2734

\bibitem[\protect\citeauthoryear{Puzia et~al.,}{Puzia et~al.}{2004}]{puzia2004vlt}
Puzia T.~H.,  et~al., 2004, A\&A, 415, 123

\bibitem[\protect\citeauthoryear{Rhode, Zepf, Kundu  \& Larner}{Rhode et~al.}{2007}]{rhode2007global}
Rhode K.~L.,  Zepf S.~E.,  Kundu A.,   Larner A.~N.,  2007, AJ, 134, 1403

\bibitem[\protect\citeauthoryear{Saito et~al.,}{Saito et~al.}{2012}]{saito2012vvv}
Saito R.~K.,  et~al., 2012, A\&As, 537, A107

\bibitem[\protect\citeauthoryear{Schlafly \& Finkbeiner}{Schlafly \& Finkbeiner}{2011}]{schlafly2011measuringData}
Schlafly E.~F.,  Finkbeiner D.~P.,  2011, AJ, 737, 103

\bibitem[\protect\citeauthoryear{Schulman, Glebbeek  \& Sills}{Schulman et~al.}{2012}]{schulman2012effect}
Schulman R.~D.,  Glebbeek v.,   Sills A.,  2012, MNRAS, 420, 651

\bibitem[\protect\citeauthoryear{Simanton-Coogan, Chandar, Miller  \& Whitmore}{Simanton-Coogan et~al.}{2017}]{simanton2017gemini}
Simanton-Coogan L.~A.,  Chandar R.,  Miller B.,   Whitmore B.~C.,  2017, AJ, 851, 63

\bibitem[\protect\citeauthoryear{Strader, Brodie, Spitler  \& Beasley}{Strader et~al.}{2006}]{strader2006globular}
Strader J.,  Brodie J.~P.,  Spitler L.,   Beasley M.~A.,  2006, AJ, 132, 2333

\bibitem[\protect\citeauthoryear{Thomas et~al.,}{Thomas et~al.}{2020}]{thomas2020hidden}
Thomas G.~F.,  et~al., 2020, AJ, 902, 89

\bibitem[\protect\citeauthoryear{Vikaeus, Zackrisson, Schaerer, Visbal, Fransson, Malhotra, Rhoads  \& Sahl{\'e}n}{Vikaeus et~al.}{2022}]{vikaeus2022conditions}
Vikaeus A.,  Zackrisson E.,  Schaerer D.,  Visbal E.,  Fransson E.,  Malhotra S.,  Rhoads J.,   Sahl{\'e}n M.,  2022, MNRAS, 512, 3030

\bibitem[\protect\citeauthoryear{Wang \& Ma}{Wang \& Ma}{2021}]{wang2021properties}
Wang S.,  Ma J.,  2021, A\&A, 649, A138

\bibitem[\protect\citeauthoryear{de Brito~Silva et~al.,}{de~Brito~Silva et~al.}{2022}]{de2022j}
de Brito~Silva D.,  et~al., 2022, A\&A, 664, A129

\makeatother
\end{thebibliography}




\appendix

\section{DECam Optical View of the Circinus galaxy}\label{app}

The observations of the Circinus galaxy were conducted using the Dark Energy Camera (DECam) instrument on the Blanco 4m telescope (Proposal ID: 2022A-572211) situated at the Cerro Tololo Inter-American Observatory in Chile. They were obtained on two consecutive half nights of June 15 and 16, 2022, using three filters (gri’), as detailed in the table \ref{table:A} below.
 The DECam is composed of 62 science 2048x4096 CCDs, offering a total
FOV of about 3 sq. degrees with a resolution of 0.2637 arcsec/pixel. We followed a 9-step random pattern to get rid of physical gaps between the CCDs in each exposure series and used the automatic pipeline for processing and stacking. Then SExtractor was used on the combined fields. Local standards were
taken from the NOAO Source Catalogue to transform our instrumental magnitudes into standard ones. 
 The data obtained from this observation was later merged with our VVVX GC candidates, which allowed us to analyse more comprehensively the properties of our cluster candidates. 
\begin{table*}
  \centering
  \caption{Observation summary in three filters (gri').}
    \begin{tabular}{llll}
    \hline
    ID   & Passband & Day I (s) & Day II (s) \\
    \hline
    1     & g     & 4480  & 6260 \\
    2     & r     & 3940  & 5900 \\
    3     & i$'$     & 4260  & 4260 \\
    \hline
    \end{tabular}%
  \label{table:A}%
\end{table*} 

\begin{figure*}
\centering
         \centering
         \includegraphics[width=\linewidth]{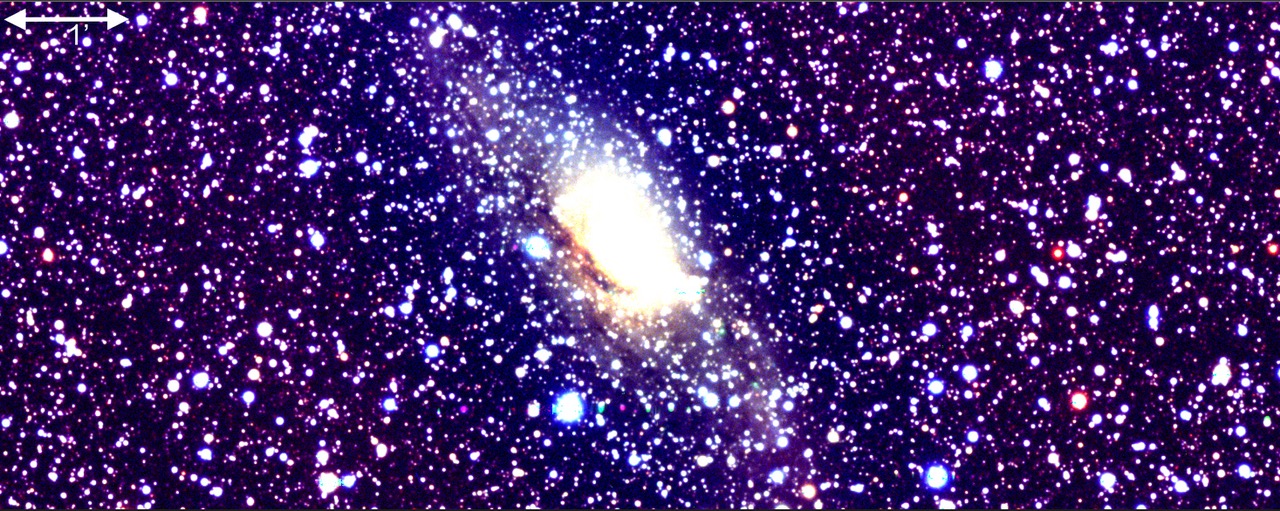}
        \caption{The composite image of the Circinus galaxy obtained through our DECam optical observation. }
        \label{A1}
\end{figure*}


\bsp	
\label{lastpage}
\end{document}